\documentclass[11pt,a4paper]{article}
\usepackage[T1]{fontenc}
\usepackage{lmodern}
\usepackage[margin=22mm]{geometry}
\usepackage{microtype}
\usepackage{amsmath}
\usepackage{graphicx}
\usepackage{booktabs,multirow}
\usepackage[font=small,labelfont=bf]{caption}
\usepackage{authblk}
\usepackage[sort&compress,super,comma]{natbib}
\usepackage[hidelinks]{hyperref}
\usepackage{xurl}
\DeclareUnicodeCharacter{00D7}{\ensuremath{\times}}
\DeclareUnicodeCharacter{00B0}{\ensuremath{{}^{\circ}}}
\DeclareUnicodeCharacter{2103}{\ensuremath{{}^{\circ}\mathrm{C}}}
\DeclareUnicodeCharacter{00B7}{\ensuremath{\cdot}}
\DeclareUnicodeCharacter{2022}{\textbullet}
\DeclareUnicodeCharacter{2032}{\ensuremath{\prime}}
\DeclareUnicodeCharacter{2212}{\ensuremath{-}}
\DeclareUnicodeCharacter{0394}{\ensuremath{\Delta}}
\DeclareUnicodeCharacter{00B5}{\ensuremath{\mu}}
\DeclareUnicodeCharacter{03BC}{\ensuremath{\mu}}
\DeclareUnicodeCharacter{03B1}{\ensuremath{\alpha}}
\DeclareUnicodeCharacter{03C1}{\ensuremath{\rho}}
\DeclareUnicodeCharacter{03BB}{\ensuremath{\lambda}}
\DeclareUnicodeCharacter{00D8}{\O}
\date{}
\title{Crystal Growth, Defect Structures and Properties of Thorium-Doped BaF\textsubscript{2} Single Crystal for Solid-State Nuclear Optical Clocks}
\author[1,2]{Lin Li}
\author[1,2]{Yuxiang Li}
\author[1]{Qiaorui Gong}
\author[1,2]{Shanming Li}
\author[1,2]{Mingzhu He}
\author[1,2]{Shuang Cai}
\author[1,2,3]{Peixiong Zhang\thanks{Corresponding author: \href{mailto:pxzhang@jnu.edu.cn}{\nolinkurl{pxzhang@jnu.edu.cn}}.}}
\author[1,2]{Chengchun Zhao\thanks{Corresponding author: \href{mailto:zhaocc205@siom.ac.cn}{\nolinkurl{zhaocc205@siom.ac.cn}}.}}
\author[1,2]{Yin Hang\thanks{Corresponding author: \href{mailto:yhang@siom.ac.cn}{\nolinkurl{yhang@siom.ac.cn}}.}}
\author[4]{Longsheng Ma}
\author[5]{Shining Zhu}
\affil[1]{Research Center of Laser Crystal, Shanghai Institute of Optics and Fine Mechanics, Chinese Academy of Sciences, Shanghai 201800, China.}
\affil[2]{Center of Materials Science and Optoelectronics Engineering, University of Chinese Academy of Sciences, Beijing 100049, China.}
\affil[3]{Department of Optoelectronic Engineering, Jinan University, Guangzhou, Guangdong 510632, China.}
\affil[4]{State Key Laboratory of Precision Spectroscopy, East China Normal University, Shanghai 200062, China.}
\affil[5]{National Laboratory of Solid-State Microstructures, School of Physics, Collaborative Innovation Center of Advanced Microstructures, Nanjing University, Nanjing, Jiangsu 210093, China.}
\hypersetup{
  pdftitle={Crystal Growth, Defect Structures and Properties of Thorium-Doped BaF2 Single Crystal for Solid-State Nuclear Optical Clocks},
  pdfauthor={Lin Li, Yuxiang Li, Qiaorui Gong, Shanming Li, Mingzhu He, Shuang Cai, Peixiong Zhang, Chengchun Zhao, Yin Hang, Longsheng Ma, Shining Zhu}
}
\begin{document}
\maketitle
\begin{abstract}
Thorium-doped fluorides are candidate host materials for solid-state nuclear clocks. Pure BaF\textsubscript{2} and \textsuperscript{232}Th-doped BaF\textsubscript{2} single crystals were grown and characterized by X-ray diffraction, Laue diffraction, Raman spectroscopy and vacuum-ultraviolet (VUV) transmission measurements. Crystal growth involved controlled preservation of an unmelted seed region and slow final cooling to limit macroscopic bubbles and improve crystal uniformity. The diffraction results support retention of the fluorite host structure in the measured samples. ICP results indicate a \textsuperscript{232}Th concentration of 2.29×10\textsuperscript{20} cm\textsuperscript{-3}, demonstrating the successful incorporation of Th into the BaF\textsubscript{2} crystals. Density functional theory (DFT) was employed to calculate charge-compensated Th monomers and 31 neutral dimer defect configurations. PBE0 calculations give band gaps of 9.379 eV for BaF\textsubscript{2} and 9.074 eV for the 90° interstitial-fluoride-compensated Th:BaF\textsubscript{2} monomer model, corresponding to nominal absorption edges of 132.2 and 136.6 nm, respectively, both shorter than the nuclear transition wavelength of 148.4 nm, supporting further evaluation of BaF\textsubscript{2} as a host for solid-state nuclear optical clocks. The pure crystal shows transmittances of 64.43\% at 148.4 nm. These results identify VUV attenuation is the main limitation of Th:BaF\textsubscript{2}, and further provide optimization recommendations regarding raw material processing, crystal growth atmosphere, and in-situ annealing, which are expected to improve VUV transparency.
\end{abstract}

\section{Introduction}\label{sec:1}

High-precision optical clocks are essential instruments for the precise measurement of time and frequency. They have irreplaceable strategic value in a broad range of applications on land, at sea, in the air and in space, including satellite navigation, spaceborne optical clocks, submarine positioning and guidance, geodesy, and optical clock networks \cite{ref1,ref2,ref3,ref4,ref5,ref6,ref7}. Among the various optical clock platforms, solid-state nuclear clocks based on nuclear transitions have attracted considerable attention because of their long-term stability and potential for miniaturization \cite{ref8,ref9,ref10,ref11,ref12}. Thorium-229 is the only known nuclide with an exceptionally low-energy nuclear excited state accessible by lasers. Its transition energy is only 8.4 eV, allowing excitation by vacuum-ultraviolet (VUV) laser light at 148.4 nm and offering a unique opportunity to develop a new generation of optical clocks based on nuclear transition frequencies \cite{ref13,ref14,ref15,ref16}. Embedding \textsuperscript{229}Th in a transparent solid with a wide band gap accommodates a large number of nuclear transition centers simultaneously and represents an important materials approach to solid-state nuclear clocks \cite{ref17,ref18,ref19,ref20}.

As the core materials of solid-state nuclear clocks, \textsuperscript{229}Th-doped crystals need to combine high doping concentrations, excellent VUV transmittance, resistance to radiation damage, and low background luminescence. Centimeter-scale and even millimeter-scale Th-doped crystals provide a materials foundation for compact nuclear clocks. Reported solid-state nuclear clock materials mainly include \textsuperscript{229}Th:CaF\textsubscript{2} \cite{ref21,ref22,ref23,ref24,ref25,ref26,ref27}, \textsuperscript{232}Th:SrF\textsubscript{2} \cite{ref28}, \textsuperscript{232}Th:LiF \cite{ref29}, \textsuperscript{232}Th:BaLiF\textsubscript{3} \cite{ref30}, \textsuperscript{229}Th:LiSrAlF\textsubscript{6} \cite{ref31,ref32}, \textsuperscript{229}Th:LiCaAlF\textsubscript{6}\cite{ref32,ref33}, \textsuperscript{232}Th:SrB\textsubscript{4}O\textsubscript{7} \cite{ref34,ref35}, and \textsuperscript{229}Th:MgF\textsubscript{2}\cite{ref14,ref36} crystals, as well as ThO\textsubscript{2}\cite{ref37} and \textsuperscript{229}ThF\textsubscript{4} \cite{ref38,ref39} films; Th:BaMgF\textsubscript{4}\cite{ref40} and Th(SO\textsubscript{4})\textsubscript{2}\cite{ref41} have also been proposed \cite{ref42}. Among these materials, \textsuperscript{232}Th-doped crystals are mainly used to investigate chemical incorporation and material properties, rather than to replace experimental verification of the \textsuperscript{229}Th nuclear transition and its self-irradiation effects. CaF\textsubscript{2} and SrF\textsubscript{2} crystals with the cubic fluorite structure exhibit high VUV transmittance, low background luminescence, and relatively mature growth techniques, and have been demonstrated to be highly promising host crystals for \textsuperscript{229}Th. Nuclear clock operation has been experimentally demonstrated in \textsuperscript{229}Th:CaF\textsubscript{2}. Using millimeter-scale \textsuperscript{229}Th:CaF\textsubscript{2} crystals, a continuous-wave VUV laser at 148.4 nm was locked to the nuclear transition to realize a solid-state nuclear clock, further demonstrating the feasibility of this system for compact nuclear clocks \cite{ref43,ref44}.

Despite the substantial advantages of current solid-state nuclear clock materials, exemplified by Th:CaF\textsubscript{2}, performance bottlenecks remain. Materials with low thermal conductivity are prone to heat accumulation under prolonged laser irradiation, which can increase the number of lattice defects. Substitution of divalent cations by Th\textsuperscript{4+} alters the local charge balance. Aggregation of Th\textsuperscript{4+}-related defects occurs in fluorite crystals, giving rise to electric field gradient (EFG) distributions associated with multiple centers and increasing spectral broadening and frequency uncertainty \cite{ref45,ref46,ref47,ref48}, thereby affecting nuclear clock accuracy. The nuclear clock transition frequency in \textsuperscript{229}Th-doped crystals is affected by the interaction between the nuclear quadrupole moment and the EFG, which produces hyperfine splitting. The EFG at the \textsuperscript{229}Th nucleus is determined by the configuration of the surrounding ions. A detailed understanding of the microscopic physical properties of \textsuperscript{229}Th-doped crystals and the development of new, high-performance fluorite hosts are therefore essential to overcoming the performance limitations of existing nuclear clocks. Currently reported charge-compensation models involving two F interstitial or a Ca vacancy cannot fully explain the experimentally observed EFG characteristics of multiple centers, including O-like, D-like centers, and high-EFG centers \cite{ref49,ref50}. Recent site-resolved laser Mössbauer studies and HAADF-STEM imaging further indicate that Th\textsuperscript{4+} ions tend to aggregate in fluorite CaF\textsubscript{2} crystals at high doping concentrations \cite{ref51}. This finding suggests that material optimization should address not only the total Th concentration but also charge compensation, the tendency of Th to aggregate, and heat treatment \cite{ref19}. Understanding the interplay between local structure, chemical bonding, and optical properties is essential for the rational optimization of optical materials. Local structure and chemical bonding influence macroscopic optical responses \cite{ref52}, motivating an atomistic investigation of charge-compensated Th defects in BaF\textsubscript{2}. Like CaF\textsubscript{2} and SrF\textsubscript{2}, BaF\textsubscript{2} has a cubic fluorite structure and shares its intrinsic advantages of a highly symmetric lattice, low phonon energy, a broad transparency range, and a low refractive index, providing a basis for studies of Th-doped crystal growth. In terms of thermal properties, BaF\textsubscript{2} has a thermal conductivity of 11.7 W/(m·K) at 300 K, higher than those of CaF\textsubscript{2} (9.71 W/(m·K)) and SrF\textsubscript{2} (8.3 W/(m·K)). Its melting point of 1368 ℃ is lower than those of CaF\textsubscript{2} (1420 ℃) and SrF\textsubscript{2} (1473 ℃), which may help reduce the volatilization of costly \textsuperscript{229}Th feedstock, although Th retention still depends on its chemical form, the atmosphere, and the holding process. Despite these favorable properties, systematic studies of bulk-grown Th:BaF\textsubscript{2} as a candidate host for solid-state nuclear clocks has not yet been explored, particularly regarding crystal growth, charge-compensating defects, and VUV transmission. The DFT calculations in this work give a band gap of 9.379 eV for pure BaF\textsubscript{2}, exceeding the \textsuperscript{229}Th nuclear transition energy. This result indicates that the ideal model has a large electronic band gap. It is necessary to investigate crystal growth, local defects, and VUV losses in Th:BaF\textsubscript{2} together.

This work presents the first systematic investigation combining crystal growth and first-principles analysis of Th-related defect structures in a \textsuperscript{232}Th:BaF\textsubscript{2} single crystal, with pure BaF\textsubscript{2} as a reference. X-ray diffraction, Laue diffraction, and Raman spectroscopy are combined to evaluate the host structure, while first-principles calculations are used to compare monomer charge-compensation mechanisms and candidate dimer configurations. VUV tranmistance spectra are further used to discuss the distinction between the electronic band gaps of ideal models and actual transmission losses. The study focuses on identifying the stability characteristics of Th-related local structures and determining the material and growth conditions that should be controlled to improve the properties of \textsuperscript{232}Th:BaF\textsubscript{2} crystals.

\section{Methods}\label{sec:2}

\subsection{Growth and Characterization}\label{sec:2.1}

Pure BaF\textsubscript{2} and Th:BaF\textsubscript{2} single crystals were successfully grown via the Bridgman method. \textsuperscript{232}Th was used as the dopant based on its chemical similarity to \textsuperscript{229}Th, along with its favorable availability and low radioactivity. Powders of \textsuperscript{232}ThF\textsubscript{4} (99.99\%), BaF\textsubscript{2} (99.99\%) and PbF\textsubscript{2} (99.99\%) were fully mixed and charged in an ultrapure graphite crucible (Ø6×110 mm). The furnace was evacuated to \textasciitilde3×10\textsuperscript{-3} Pa. The charge was melted over approximately 9 h at a temperature above the melting point of BaF\textsubscript{2}. Then the growth process was driven by lowering the crucible at a rate of 0.8 mm/h in the gradient zone. At last, the as-grown crystal was cooled to room temperature at a rate of 50\textasciitilde60 ℃/h. The grown single crystals were processed through orientation, slicing, and polishing to obtain slices with a thickness of 1 mm. Portions of the crystal samples were ground to powders for characterization.

The crystal structure was characterized using a D/max 2550 X-ray diffractometer (XRD) with Cu K\textsubscript{α} radiation. The Th doping concentration was determined by inductively coupled plasma optical emission spectrometry (ICP-OES) using an Agilent 5800 device at a radio frequency power of 1.2 kW. Crystal orientation and crystalline quality were assessed by X-ray Laue diffraction using a single-crystal orientation system operated at a tube voltage of 30 kV and a tube current of 20 mA. Laue patterns were collected at different positions on crystal with an exposure time of 120 s per pattern. Raman spectra are recorded via a continuous-wave laser of 532 nm. Prior to Raman measurements on the samples, standard silicon wafers were used to perform peak position calibration. The estimated uncertainty in peak position is 0.2 cm\textsuperscript{-1}. The VUV transmission spectra in 120-200 nm were obtained using the VUV/UV spectrophotometer (ML 6500) of the Hanover Laser Center.

\subsection{Calculation details}\label{sec:2.2}

All first-principles calculations were performed using CASTEP within the plane-wave pseudopotential framework \cite{ref53}. For geometry optimization, the Perdew--Burke--Ernzerhof (PBE) exchange-correlation functional within the generalized gradient approximation (GGA) was used \cite{ref54,ref55,ref56,ref57}. OTFG Ultra-soft pseudopotentials are adopted for all atoms. The valence electronic configurations for pseudo atoms Ba, F, and Th are 5s\textsuperscript{2} 5p\textsuperscript{6} 6s\textsuperscript{2}, 2s\textsuperscript{2} 2p\textsuperscript{5} and 6s\textsuperscript{2} 6p\textsuperscript{6} 7s\textsuperscript{2} 6d\textsuperscript{2}, respectively. After convergence testing, the plane-wave cut-off energy is set to 500 eV for all self-consistent field relaxation. All the models were optimized fully by relaxing the cell parameters and ionic positions. A \emph{k}-grid with separation of 0.07/Å was used in the Brillouin zone integration to ensure the relaxation accuracy of 5×10\textsuperscript{-7} eV/atom, 0.01 eV/Å, 0.02 GPa, and 0.0005 Å in a 2×2×2 supercell, respectively. The PBE0 hybrid functional \cite{ref58} was used to calculate the electronic properties, with approximate \emph{k}-points separation of~0.01/Å for band structure calculation and 0.04/Å for density of states (DOS) calculations \cite{ref59}. The crystal structure was obtained from the Crystallography Open Database and analyzed via VESTA \cite{ref60,ref61}.

\subsection{Defect energy}\label{sec:2.3}

All defect models were constructed from a 2 × 2 × 2 supercell of the conventional BaF\textsubscript{2} unit cell containing 32 Ba atoms and 64 F atoms. A Th atom substitutes for a Ba atom on a lattice site and is denoted as Th\textsubscript{Ba}\textsuperscript{••}. For a Th monomer, two charge-compensation mechanisms are considered, a Ba vacancy, V\textsubscript{Ba}′′, and F interstitials, Fi, corresponding to the local defect configurations {[}Th\textsubscript{Ba}\textsuperscript{••}-V\textsubscript{Ba}′′{]} and {[}Th\textsubscript{Ba}\textsuperscript{••}-2F\textsubscript{i}′{]}, respectively. For the Th dimer calculations, two Th atoms substitute for Ba atoms on neighboring sites, with four F atoms providing charge compensation to form a {[}2Th\textsubscript{Ba}\textsuperscript{••}-4F\textsubscript{i}′{]} defect cluster; the supercell composition is Th\textsubscript{2}Ba\textsubscript{30}F\textsubscript{68}.

The defect formation energies of the Th monomer configurations {[}Th\textsubscript{Ba}\textsuperscript{••}-2F\textsubscript{i}′{]} and {[}Th\textsubscript{Ba}\textsuperscript{••}-V\textsubscript{Ba}\textsuperscript{′′}{]} in the supercell are calculated using the following equations, respectively \cite{ref62} :

\begin{equation}
\label{eq:1}
E_{\mathrm f} = E_{\mathrm d}-E_0-\mu_{\mathrm{Th}}-2\mu_{\mathrm F}+\mu_{\mathrm{Ba}}
\end{equation}

\begin{equation}
\label{eq:2}
E_{\mathrm f} = E_{\mathrm d}-E_0-\mu_{\mathrm{Th}}+2\mu_{\mathrm{Ba}}
\end{equation}

where \emph{E\textsubscript{d}} is the total energy of the Th:BaF\textsubscript{2} containing defects, \emph{E\textsubscript{0}} is the energy of the pure BaF\textsubscript{2} supercell, \emph{µ\textsubscript{Th, }}\emph{µ\textsubscript{F,} µ\textsubscript{Ba}} are the chemical potentials of Th, F, and Ba. Under equilibrium conditions, the chemical potentials satisfy:

\begin{equation}
\label{eq:3}
\mu_{\mathrm{BaF_2}} = \mu_{\mathrm{Ba}}+2\mu_{\mathrm F},\quad \mu_{\mathrm{Ba}}\leq\mu_{\mathrm{Ba\text{-}bulk}},\quad\mu_{\mathrm F}\leq\mu_{\mathrm{F\text{-}gas}}
\end{equation}

The heat of formation Δ of BaF\textsubscript{2} is negative, and can be expressed as:

\begin{equation}
\label{eq:4}
\Delta=\mu_{\mathrm{BaF_2\text{-}bulk}}-\mu_{\mathrm{Ba}}-2\mu_{\mathrm F}
\end{equation}

The heat of formation satisfied the following equation:

\begin{equation}
\label{eq:5}
\mu_{\mathrm{Ba\text{-}bulk}}+\Delta\leq\mu_{\mathrm{Ba}}\leq\mu_{\mathrm{Ba\text{-}bulk}}
\end{equation}

\begin{equation}
\label{eq:6}
\mu_{\mathrm{F\text{-}gas}}+\Delta\leq\mu_{\mathrm F}\leq\mu_{\mathrm{F\text{-}gas}}
\end{equation}

Then, the defect formation energies for {[}Th\textsubscript{Ba}\textsuperscript{••}-2F\textsubscript{i}′{]} and {[}Th\textsubscript{Ba}\textsuperscript{••}-V\textsubscript{Ba}′′{]} can be expressed as follows:

\begin{equation}
\label{eq:7}
E_{\mathrm f\,(\mu_{\mathrm{Ba}}=\mu_{\mathrm{Ba\text{-}bulk}},\,\mu_{\mathrm F}=\mu_{\mathrm{F\text{-}gas}})}\leq E_{\mathrm f}\leq E_{\mathrm f\,(\mu_{\mathrm{Ba}}=\mu_{\mathrm{Ba\text{-}bulk}},\,\mu_{\mathrm F}=\mu_{\mathrm{F\text{-}gas}})}-\Delta
\end{equation}

\begin{equation}
\label{eq:8}
E_{\mathrm f\,(\mu_{\mathrm{Ba}}=\mu_{\mathrm{Ba\text{-}bulk}})}+2\Delta\leq E_{\mathrm f}\leq E_{\mathrm f\,(\mu_{\mathrm{Ba}}=\mu_{\mathrm{Ba\text{-}bulk}})}
\end{equation}

The defect-cluster binding energy (\emph{ΔE \textsubscript{cluster}}) of a supercell containing a Th dimer {[}2Th\textsubscript{Ba}\textsuperscript{••}-4F\textsubscript{i}′{]} \cite{ref63,ref64} can be used to evaluate the aggregation tendency of Th\textsuperscript{4+} ions under the same conditions:

\begin{equation}
\label{eq:9}
\Delta E_{\mathrm{cluster}}=E_{\mathrm{tot}}+(m+n+w-1)E_0-mE_1-nE_2-wE_3
\end{equation}

where~\emph{E}\textsubscript{0}\hspace{0pt}~is the total energy of the~2×2×2 BaF\textsubscript{2} supercell, \emph{E}\textsubscript{1}\hspace{0pt}, \emph{E}\textsubscript{2}\hspace{0pt}, and~\emph{E}\textsubscript{3}\hspace{0pt}~are the total energies of the isolated reference defects Th\textsubscript{Ba}\textsuperscript{••}, V\textsubscript{Ba}\textsuperscript{′′}, and F\textsubscript{i}, respectively. For Ba\textsubscript{30}Th\textsubscript{2}F\textsubscript{68}, m = 2, n = 0 and w = 4. All dimer configurations considered here are charge-neutral and have identical compositions, allowing their relative energies to be compared using a common reference. This quantity serves as a relative binding energy indicator to compare the energetic favorability among different cluster configurations, rather than an absolute thermodynamic formation energy.

\section{Results and discussion}\label{sec:3}

\subsection{Crystal growth and properties}\label{sec:3.1}

\begin{figure}[!htbp]
\centering
\includegraphics[width=\linewidth]{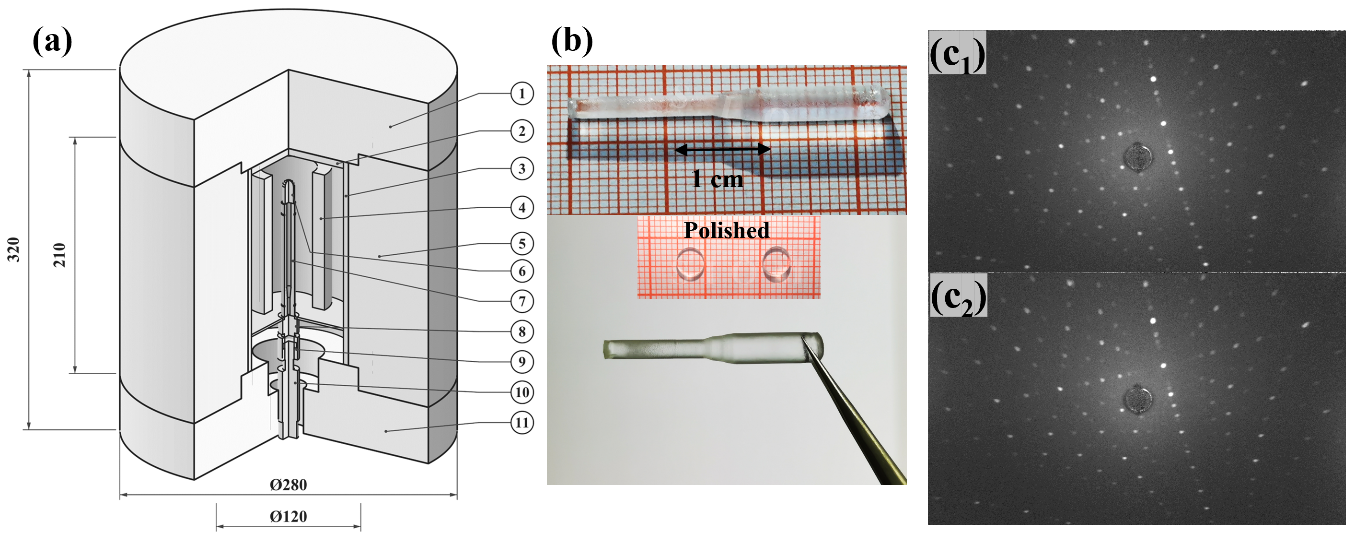}
\caption{(a) Bridgman-method crystal growth furnace chamber: 1 Upper rigid graphite felt, 2 Graphite cover plate, 3 Graphite side cylinder, 4 Graphite heating element, 5 Soft graphite felt insulation, 6 Graphite crucible lid, 7 Graphite crucible, 8 Graphite lower crucible cover, 9 Graphite crucible seat, 10 Graphite support rod, 11 Lower rigid graphite felt; (b) As-grown \textsuperscript{232}Th-doped BaF\textsubscript{2} crystal and two polished plates cut from the same crystal; (c) X-ray Laue diffraction patterns at two different positions along the crystal.}\label{fig:1}
\end{figure}

The \textsuperscript{232}Th:BaF\textsubscript{2} crystal shown in Figure~\ref{fig:1}(b) was grown by the Bridgman method described in Section~\ref{sec:2.1}, using the furnace illustrated in Figure~\ref{fig:1}(a). BaF\textsubscript{2} and ThF\textsubscript{4} were used as raw materials for the growth of the \textsuperscript{232}Th:BaF\textsubscript{2} crystal, with a small amount of PbF\textsubscript{2} added as an oxygen scavenger. The system was evacuated to \textasciitilde3×10\textsuperscript{-3} Pa and backfilled with an inert gas. Melting, crystal growth, and cooling processes were controlled by adjusting the heating power in stages. The main body of the as-grown \textsuperscript{232}Th:BaF\textsubscript{2} crystal was relatively transparent, as shown in Figure~\ref{fig:1}(b). Figure~\ref{fig:1}(c) shows the X-ray Laue diffraction patterns collected at different positions along the same \textsuperscript{232}Th:BaF\textsubscript{2} crystal. Both patterns exhibit discrete, regularly arranged diffraction spots, indicating good single-crystal quality in the measured regions. Based on the Laue characterization, two transparent plates, each 1 mm thick, were cut from selected regions of the same crystal and optically polished, as shown in Figure~\ref{fig:1}(b). The crucible geometry, axial temperature distribution, staged cooling schedule, and residual solid layer at the crucible bottom were jointly controlled to limit bubble formation and associated scattering defects.

\begin{figure}[!htbp]
\centering
\includegraphics[width=\linewidth]{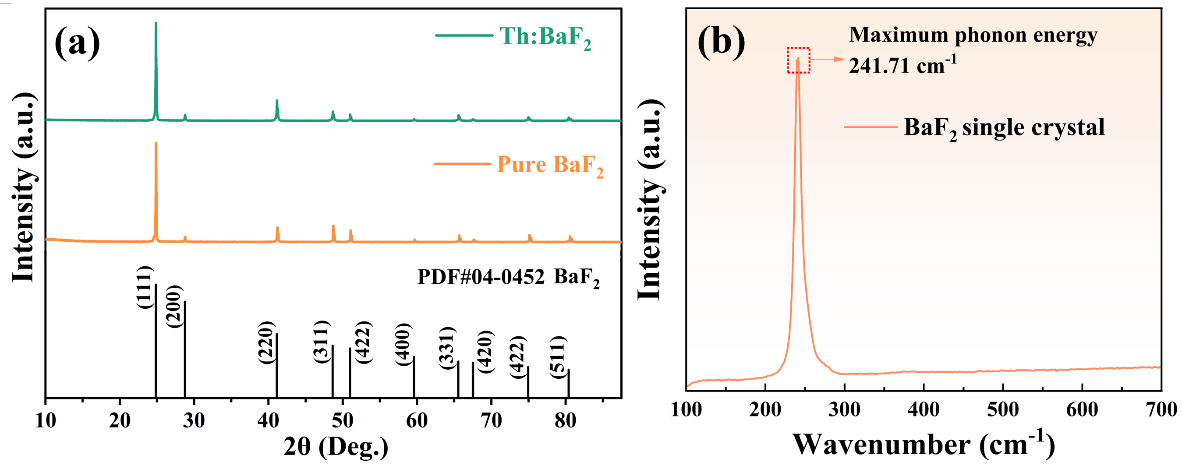}
\caption{(a) XRD patterns of pure BaF\textsubscript{2} and Th-doped BaF\textsubscript{2} crystals compared with the standard diffraction, (b) Raman spectrum of the pure BaF\textsubscript{2}.}\label{fig:2}
\end{figure}

Figure~\ref{fig:2}(a) shows the XRD patterns of the as-grown pure BaF\textsubscript{2} and \textsuperscript{232}Th:BaF\textsubscript{2} crystals. All peak positions agree well with the standard powder diffraction file (PDF\#04-0452), indicating that the cubic BaF\textsubscript{2} structure is retained after \textsuperscript{232}Th doping. Both sets of samples exhibit relatively sharp diffraction peaks, indicating good crystallinity. No distinct additional diffraction peaks from impurity phases are observed within the measured range, suggesting that \textsuperscript{232}Th doping does not induce a transformation of the host phase or appreciable precipitation of a crystalline secondary phase. As shown in Figure~\ref{fig:2}(b), pure BaF\textsubscript{2} has a dominant Raman peak near 241.71 cm\textsuperscript{-1}, which can be assigned to the T\textsubscript{2g} vibrational mode of the cubic fluorite structure. This agrees with the characteristic peak at approximately 241 cm\textsuperscript{-1} reported in the literature and is lower than those of other fluoride crystals, such as CaF\textsubscript{2} (321.5 cm\textsuperscript{-1}) \cite{ref65} and SrF\textsubscript{2} (285.0 cm\textsuperscript{-1}) \cite{ref66}. Defects, doping, or the growth environment may slightly reduce the vibrational frequency but do not change the intrinsic maximum phonon energy. This comparison indicates that the lattice-vibration energy quanta in BaF\textsubscript{2} are smaller, which is favorable for reducing nuclear-level broadening after Th doping and may enable the development of solid-state nuclear clocks with higher precision.

A high Th doping concentration enhances the nuclear transition signal and is a key factor in developing solid-state nuclear optical clocks. Table~\ref{tab:1} lists the \textsuperscript{232}Th content and corresponding doping concentration in the grown \textsuperscript{232}Th:BaF\textsubscript{2} crystal. \textsuperscript{232}Th ions were successfully incorporated into the BaF\textsubscript{2} lattice. The doping concentration at the top of the crystal is 2.29×10\textsuperscript{20} cm\textsuperscript{-3}, exceeding the initial concentration in the starting materials. The maximum segregation coefficient (\emph{k\textsubscript{eff}}) is 1.38, which is also higher than \textsuperscript{232}Th:SrF\textsubscript{2} crystal (1.14) \cite{ref28}.

\begin{table}[!htbp]
\centering
\caption{Thorium content in initial ingredients and the as-grown Th:BaF\textsubscript{2} crystal.}\label{tab:1}
 \begin{tabular}{@{}lrrc@{}}
\toprule
Samples & Initial ingredients & Th:BaF\textsubscript{2} (top) & $k_{\mathrm{eff}}$ (max) \\
\midrule
Weight percentage (wt.\ \%) & 1.797 & 2.482 & \multirow{2}{*}{1.38} \\
Doping concentration ($\times10^{20}$ cm$^{-3}$) & 1.658 & 2.290 & \\
\bottomrule
\end{tabular}
\end{table}

\subsection{Crystal structure}\label{sec:3.2}

\subsubsection{Defect and band structure}\label{sec:3.2.1}

\begin{figure}[!htbp]
\centering
\includegraphics[width=\linewidth]{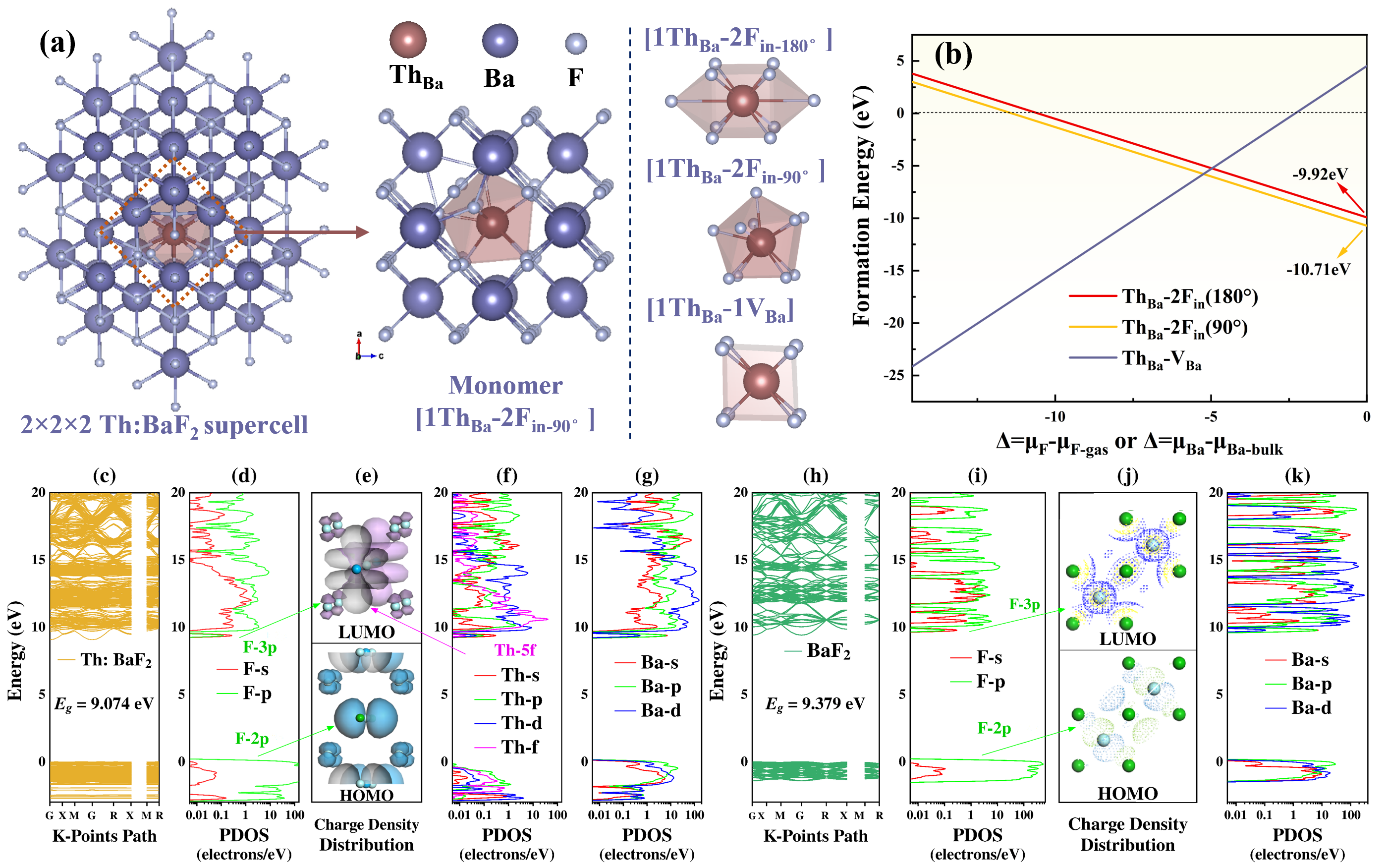}
\caption{(a) Thermodynamically stable 2 × 2 × 2 supercell of Th:BaF\textsubscript{2} and Th\textsuperscript{4+} monomer structures; (b) Formation energy of each defect configurations at different \emph{μ\textsubscript{Ba}} or \emph{μ\textsubscript{F}}; The band structure, PDOS, and HOMO-LUMO of per elements in Th:BaF\textsubscript{2} crystal (c)-(g) and BaF\textsubscript{2} crystal (h)-(k).}\label{fig:3}
\end{figure}

As shown in Figure~\ref{fig:3}(a) BaF\textsubscript{2} is fluorite structured with cubic \ensuremath{Fm\bar{3}m} space group. Ba\textsuperscript{2+} cations are coordinated by 8 equivalent F\textsuperscript{-} anions in a body-centered cubic coordination, with a uniform Ba-F bond distance of 2.70 Å. Substitution of Ba\textsuperscript{2+} by Th\textsuperscript{4+} introduces a positive effective charge, which should be compensated by defects with negative effective charges. Figures~\ref{fig:3}(a) compares the 90° and 180° configurations compensated by two F\textsubscript{i} interstitials and the configuration compensated by V\textsubscript{Ba}\textsuperscript{′′}. Figure~\ref{fig:3}(b) presents the thermodynamic stability of the Th monomer structures obtained from DFT calculations, with the formation energies of the three configurations evaluated at different \emph{µ\textsubscript{F}} and \ensuremath{\mu_{\mathrm{Ba}}} values. Table~\ref{tab:2} lists the \emph{E\textsubscript{f}} values at Δ=0 and compares the structural distortion parameters of the pure and Th-doped BaF2 models and the experimental lattice parameters. A more negative formation energy indicates that formation of the defect is more thermodynamically favorable. However, a negative formation energy does not imply that defects can form without limit. During actual growth, the chemical potentials of all elements cannot readily reach their elemental reference energies simultaneously. Consequently, the actual formation energies of F\textsubscript{i} defects will be higher than those listed in Table 2, whereas those of V\textsubscript{Ba}′′ will be correspondingly lower. As in Th:CaF\textsubscript{2} and Th:SrF\textsubscript{2}, the formation energies of the F\textsubscript{i} interstitial defect configurations in Th:BaF\textsubscript{2} are negative and substantially lower than those of the cation-vacancy configuration. The 90° F\textsubscript{i} configuration has a lower formation energy than the 180° F\textsubscript{i} configuration, indicating that formation of the 90° F\textsubscript{i} interstitial configuration is also thermodynamically favored in Th:BaF\textsubscript{2}. Furthermore, both thermodynamic and kinetic factors should be considered when assessing the ease of defect formation. In Table~\ref{tab:2}, the changes in the lattice parameters and cell volume of the 90° {[}Th\textsubscript{Ba}\textsuperscript{••}-2F\textsubscript{i}′{]} configuration in Th:BaF\textsubscript{2} are smaller than those of the 180° {[}ThBa\textsuperscript{••}-2F\textsubscript{i}′{]} configuration, indicating less distortion. Thus, both kinetic and thermodynamic factors favor formation of Th monomer defects in the 90° {[}Th\textsubscript{Ba}\textsuperscript{••}-2F\textsubscript{i}′{]} configuration in Th:BaF\textsubscript{2}.

As shown in Figures~\ref{fig:3}(c-k), Th doping slightly reduces the band gap. The PDOS indicate that Th-related unoccupied states contribute to the electronic structure near the conduction-band edge. The impurity levels introduced by Th doping reduce the band gaps slightly below the 9.379 eV value of intrinsic BaF\textsubscript{2} and are expected to redshift the ultraviolet cutoff edge of Th:BaF\textsubscript{2}. Using \emph{λ} = hc/E, the band gaps calculated with the PBE0 hybrid functional for pure BaF\textsubscript{2} (9.379 eV) and Th:BaF\textsubscript{2} (9.074 eV) correspond to approximately 132.2 nm and 136.6 nm, respectively. Although Th incorporation produces an estimated redshift of 4.4 nm, both edges remain at wavelengths shorter than 148.38 and 150 nm. These results suggest that the nuclear transition energy requirement may be met in theory. However, actual optical absorption is also influenced by transition matrix elements, excitonic effects, defect states, impurities, surface conditions, and scattering.

\begin{table}[!htbp]
\centering
\caption{Structural parameters in BaF\textsubscript{2} and Th-doped BaF\textsubscript{2} crystals from DFT calculation and crystal growth.}\label{tab:2}
\resizebox{\linewidth}{!}{%
\begin{tabular}{@{}llccrrrr@{}}
\toprule
 & Structure & \shortstack{Lattice parameters\\$a/b/c$ (Å)} & $\Delta a/\Delta b/\Delta c$ (Å) & $V$ (Å$^3$) & $\Delta V$ (Å$^3$) & $\rho$ (g/cm$^3$) & $\Delta\rho$ (g/cm$^3$) \\
\midrule
\multirow{4}{*}{\shortstack[l]{DFT\\calculation}}
 & Pure BaF\textsubscript{2} & 6.324 & -- & 252.87 & -- & 4.605 & -- \\
 & [Th\textsubscript{Ba}-2F\textsubscript{in1}] & 6.315/6.315/6.328 & $-0.009/-0.009/+0.004$ & 252.04 & $-0.83$ & 4.739 & 0.134 \\
 & [Th\textsubscript{Ba}-2F\textsubscript{in2}] & 6.318/6.323/6.318 & $-0.006/-0.001/-0.006$ & 252.40 & $-0.47$ & 4.723 & 0.118 \\
 & [Th\textsubscript{Ba}-V\textsubscript{Ba}] & 6.329/6.324/6.329 & 0.005/0/0.005 & 253.29 & 0.42 & 4.563 & $-0.043$ \\
\addlinespace
\multirow{3}{*}{\shortstack[l]{Crystal\\growth}}
 & Standard BaF\textsubscript{2} & 6.200 & -- & 238.34 & -- & 4.886 & -- \\
 & Pure BaF\textsubscript{2} & 6.119 & $-0.081$ & 238.26 & $-0.08$ & 4.887 & 0.001 \\
 & \textsuperscript{232}Th:BaF\textsubscript{2} & 6.189 & $-0.011$ & 237.15 & $-1.19$ & 4.910 & 0.023 \\
\bottomrule
\end{tabular}
}
\end{table}

\subsubsection{Th dimer clusters}\label{sec:3.2.2}

\begin{figure}[!htbp]
\centering
\includegraphics[width=\linewidth]{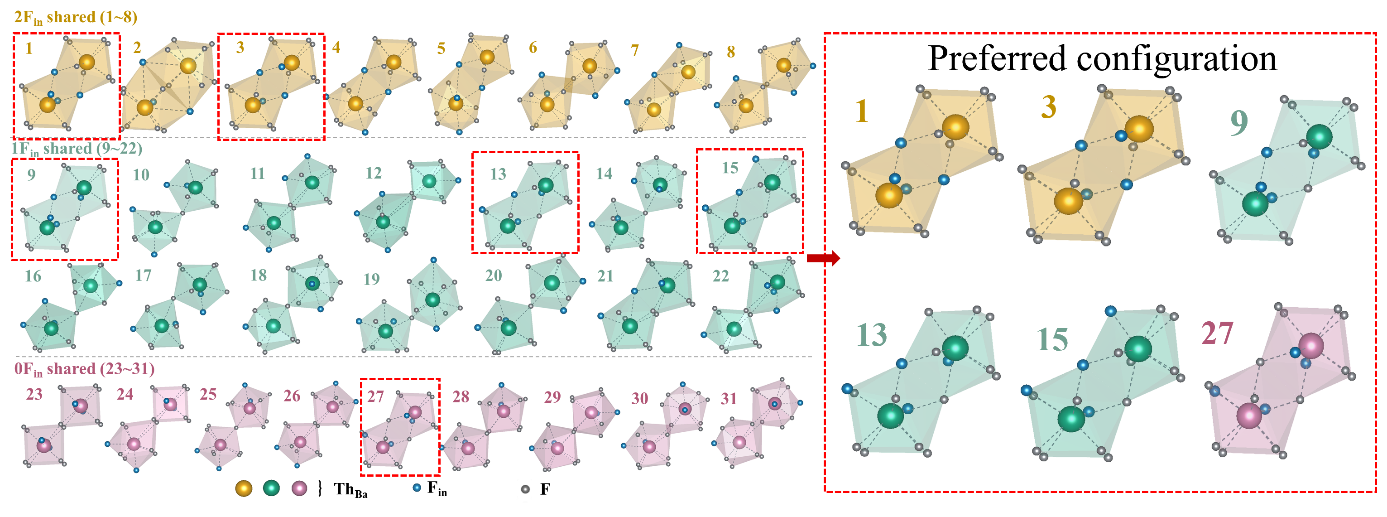}
\caption{Thermodynamically dimer clusters and preferred configurations of Th\textsuperscript{4+} in BaF\textsubscript{2}.}\label{fig:4}
\end{figure}

\begin{figure}[!htbp]
\centering
\includegraphics[width=\linewidth]{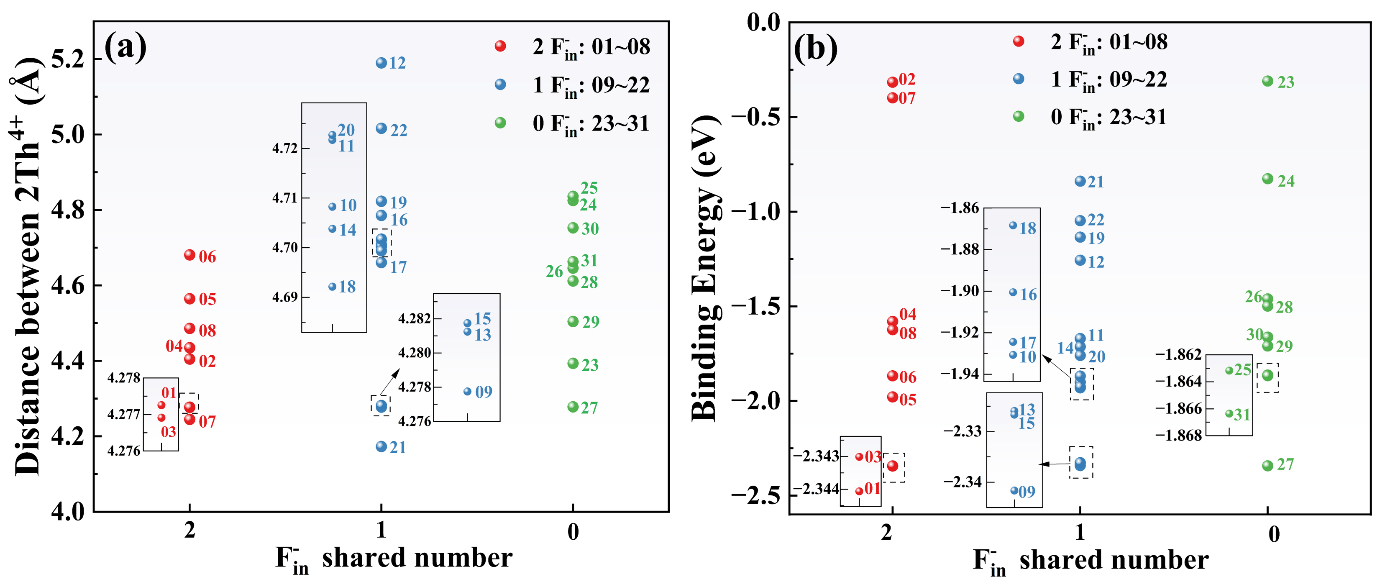}
\caption{Dimer clusters of Th\textsuperscript{4+} in BaF\textsubscript{2} (a) distance between two Th\textsuperscript{4+} ions, (b) defect cluster binding energy.}\label{fig:5}
\end{figure}

Based on the F\textsubscript{i}\textsuperscript{-}-compensated Th monomers, 31 initial charge-neutral {[}2Th\textsubscript{Ba}\textsuperscript{••}--4F\textsubscript{i}\textsuperscript{′}{]} dimer configurations were constructed to analyze local correlations between dopant centers. These 31 charge-neutral Th dimer defect configurations were generated using different nearest-neighbor F\textsubscript{i}\textsuperscript{-} positions. According to the number of shared F\textsubscript{i}\textsuperscript{-} ions in the initial structures, they are divided into three groups: 2F\textsubscript{i}-shared (1-8), 1F\textsubscript{i}-shared (9-22), and 0F\textsubscript{i}-shared (23-31). Figure~\ref{fig:4} shows the thermodynamically stable Th dimer structures in the Th:BaF\textsubscript{2} supercell. Configurations 1, 3, 9, 13, 15, and 27 were selected as the preferred configurations. These 6 configurations have similar Th arrangements and local F coordination frameworks and exhibit relatively symmetric overall structures. Figure~\ref{fig:5}(a) presents the interionic distances and defect-cluster binding energies of the Th dimers. With the binding energy defined relative to two separated, charge-neutral Th monomer defects, all 31 dimer configurations have negative binding energies, indicating an energetic advantage for dimer formation. In particular, the binding energies of configurations 1, 3, 9, 13, 15, and 27 fall within approximately −2.34 to −2.32 eV, lower than those of the remaining configurations, with only small differences among them. As shown in Figure~\ref{fig:5}(b), different distributions of F\textsubscript{i}\textsuperscript{-} ions and surrounding coordination environments still produce appreciable energy differences at similar Th--Th distances. The relaxed Th--Th distances of the 31 configurations range from \textasciitilde4.17 Å to 5.19 Å, whereas those of the six preferred configurations are all concentrated near 4.28 Å, further demonstrating the similarity of their local geometries. Overall, the six preferred configurations share similar geometries, relatively high structural symmetry, low cluster binding energies, and similarly short Th-Th distances, providing clues to defect structures that may occur in actual crystals.

\subsection{VUV transmission and defect control}\label{sec:3.3}

\begin{figure}[!htbp]
\centering
\includegraphics[width=\linewidth]{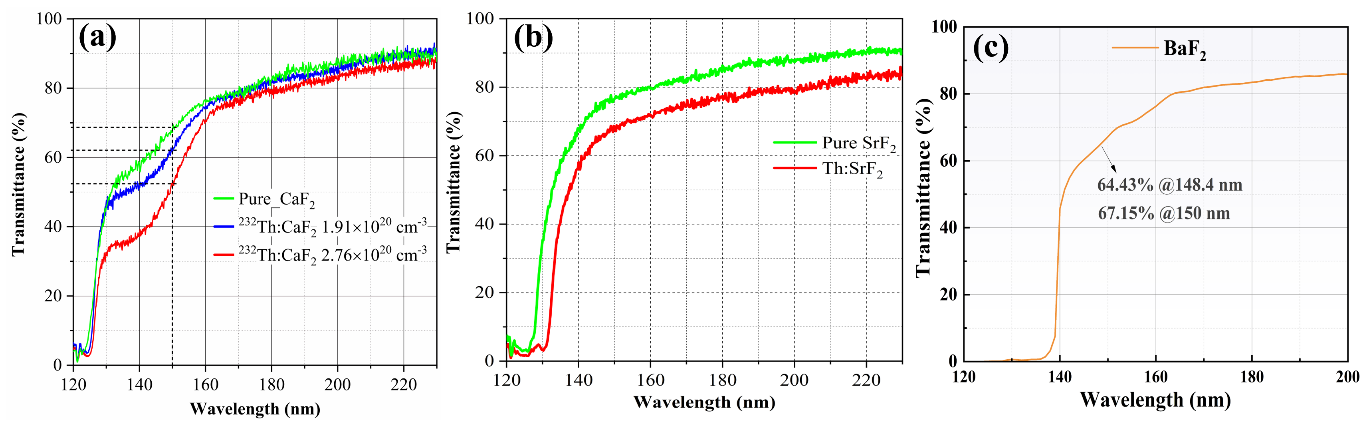}
\caption{VUV transmittances of (a) pure and \textsuperscript{232}Th-doped CaF\textsubscript{2} \cite{ref22}, (b) pure and \textsuperscript{232}Th-doped SrF\textsubscript{2} \cite{ref28}, and (c) the pure BaF\textsubscript{2} grown in this work. (1mm thickness)}\label{fig:6}
\end{figure}

Figure~\ref{fig:6} compares the VUV transmission spectra of reported pure and \textsuperscript{232}Th-doped CaF\textsubscript{2} and SrF\textsubscript{2} and as-grown BaF\textsubscript{2} crystals. The transmittance of the pure BaF\textsubscript{2} sample rises rapidly near \textasciitilde140 nm, reaching 64.43\% at 148.4 nm and 67.15\% at 150 nm. Although slightly lower than CaF\textsubscript{2} and SrF\textsubscript{2}, these values indicate good transmittance near the nuclear transition wavelength. The measured transmittance includes the combined effects of interface reflection, bulk absorption, and scattering. When additional surface absorption and scattering are negligible or have been corrected separately, internal scattering is weak, and detection is dominated by collimated transmission, the relationship between these factors can be approximated by the following expression, neglecting multiple internal reflections \cite{ref67}:

\begin{equation}
\label{eq:10}
T(\lambda)\approx(1-R_1)(1-R_2)\exp\big\{-\big[\alpha_{\mathrm{abs}}(\lambda)+\alpha_{\mathrm s}(\lambda)\big]L\big\}
\end{equation}

where \emph{R\textsubscript{1}} and \emph{R\textsubscript{2}} are the reflectances of the entrance and exit surfaces, respectively; \emph{L} is the geometric propagation length of light inside the crystal and is approximately equal to the sample thickness at near-normal incidence; \emph{α\textsubscript{abs}} is the bulk absorption coefficient; and \emph{α\textsubscript{s}} is the effective scattering-loss coefficient associated with the detection acceptance conditions. All optical parameters may vary with wavelength. The band gaps calculated in Section~\ref{sec:3.2} describe the electronic properties of periodic structural models, whereas actual crystals may also contain impurities, nonstoichiometric defects, and microscopic inhomogeneous regions that affect their transmittance. Charge-compensating defects and their associated structures formed after Th\textsuperscript{4+} substitution may influence VUV absorption by modifying local electronic states.

The VUV transmission spectra of 1-mm-thick plate cut from the \textsuperscript{232}Th:BaF\textsubscript{2} crystal were further measured. The transmittances at 150 nm were \textasciitilde14.44\%. In Na\textsubscript{2}CeF\textsubscript{6}, minor Ce\textsuperscript{3+} impurities have been shown to degrade optical performance, highlighting the importance of controlling impurity formation during fluoride crystal preparation \cite{ref68}. Shendrik, R. \emph{et al.}\cite{ref69} reported an absorption band at 8.4 eV (147.6 nm) associated with oxygen-impurity-related oxygen--fluorine-vacancy complex centers in BaF\textsubscript{2}, close to the nuclear transition wavelength. Therefore, fluorination of the raw materials and careful control of residual oxygen and moisture during Th:BaF\textsubscript{2} growth may help reduce losses associated with oxygen contamination. Ogawa \emph{et al.}\cite{ref70} studied F vacancies, V\textsubscript{F}\textsuperscript{′}, and F\textsubscript{i} interstitial defects in fluorite AF\textsubscript{2} compounds (A = Ca (1.12 Å), Sr (1.26 Å), Ba (1.42 Å), etc.). They found that the formation energy of F\textsubscript{i} decreases with increasing ionic radius, whereas that of V\textsubscript{F} follows the opposite trend. The former is mainly related to the available interstitial space, while the latter is also affected by local bonding. This indicates that BaF\textsubscript{2}, with its larger lattice space, more readily accommodates F\textsubscript{i}. The V\textsubscript{F} migration barrier in BaF\textsubscript{2} is 0.38 eV, which is 0.11 eV higher than the 0.27 eV value in CaF\textsubscript{2}; the F\textsubscript{i} interstitial migration barrier is 0.55 eV in both BaF\textsubscript{2} and CaF\textsubscript{2}. These results provide a reference for understanding the dynamics of the anion sublattice in BaF\textsubscript{2}. From the perspective of defect kinetics, the temperature programs during the final growth and cooling stages and in situ annealing may therefore also affect the final distribution of F\textsubscript{i} related defects. According to Beeks \emph{et al.} \cite{ref71}, thermal dissociation during growth under vacuum leads to F depletion in \textsuperscript{229}Th:CaF\textsubscript{2}, while reactions between CaF\textsubscript{2} and residual water produce nonstoichiometric or F deficient crystals. Crystal growth is a dynamic process in which defects continuously accumulate, accelerating radiation damage. F deficiency results in the formation of metallic Ca nanoparticles that absorb and scatter VUV light, reducing the transmittance. Similar scattering may also occur in BaF\textsubscript{2}. In addition, the size mismatch between the dopant Th\textsuperscript{4+} (1.05 Å) and the host Ba\textsuperscript{2+} (1.42 Å) may increase the number of mobile defects and thereby contribute to optical losses. To obtain Th:BaF\textsubscript{2} crystals with higher transmittance and greater radiation resistance, it is therefore necessary to focus on reducing internal defects. Alongside improvements in raw material purity and fluorination, in situ annealing can be further optimized by considering inert or fluorine-containing atmospheres, together with growth under F-rich conditions, to improve VUV transparency of the crystals.

Table~\ref{tab:3} summarizes several fundamental properties of current Th-doped crystalline candidate materials for solid-state nuclear clocks. BaF\textsubscript{2} offers a relatively simple host structure and demonstrated Th incorporation, although its doped samples currently suffer from substantial VUV losses. Compared with Th-doped SrF\textsubscript{2} and CaF\textsubscript{2}, which also have the cubic fluorite structure, Th:BaF\textsubscript{2} has a lower melting point \cite{ref72} and is easier to grow. At present, \textsuperscript{229}Th feedstock is scarce and expensive, and is almost exclusively available at Oak Ridge National Laboratory in the United States, through the decay of \textsuperscript{233}U (1.56×10\textsuperscript{5} y). The actual retention of \textsuperscript{229}Th feedstock depends jointly on the chemical form of the Th-containing components, the atmosphere, and the holding process, but a lower melting point may help reduce \textsuperscript{229}Th volatilization. Compared with LiF and BaLiF\textsubscript{3} crystals, Th:BaF\textsubscript{2} can achieve Th doping concentrations approximately two orders of magnitude higher. Our previous study \cite{ref22} noted a weak negative correlation between VUV transmittance and Th doping concentration. Compared with LiSrAlF\textsubscript{6} (\ensuremath{P\bar{3}1c}), MgF\textsubscript{2} (\ensuremath{P4_2/mnm}), and SrB\textsubscript{4}O\textsubscript{7} (\ensuremath{Pmn2_1}), the simple cubic structure of BaF\textsubscript{2} facilitates the preparation of high-quality crystals. Changes in the local lattice environment affect nuclear-level broadening on the kHz scale and nuclear-level frequency shifts on the GHz scale, with nuclear-level broadening being an important cause of reduced nuclear clock precision \cite{ref12}. Therefore, efforts to increase the number of effective nuclear transition centers should be accompanied by joint evaluation of losses at 148.4 nm, background luminescence, temperature response, and radiation stability.

\begin{table}[!htbp]
\centering
\caption{Properties of Th-doped crystals for Solid-state Nuclear Optical Clock.}\label{tab:3}
\resizebox{\linewidth}{!}{%
\begin{tabular}{@{}lcc l cccl@{}}
\toprule
\shortstack{Host\\crystal} & \shortstack{Space\\group} & \shortstack{Melt point\\($^\circ$C)} &
\shortstack{Th doping\\(cm$^{-3}$)} & \shortstack{Band gap\\(eV)} & \shortstack{Coordination\\number} & \shortstack{Cationic\\radius (Å)} & Ref. \\
\midrule
BaF\textsubscript{2} & $Fm\bar{3}m$ & 1368 & \textsuperscript{232}Th: $\sim2.29\times10^{20}$ & 9.074 & 8 & 1.42 & This work \\
SrF\textsubscript{2} & $Fm\bar{3}m$ & 1473 & \textsuperscript{232}Th: $\sim6.29\times10^{20}$ & \multirow{2}{*}{8.941} & 8 & 1.26 & \cite{ref28} \\
\multirow{2}{*}{CaF\textsubscript{2}} & \multirow{2}{*}{$Fm\bar{3}m$} & \multirow{2}{*}{1420} & \textsuperscript{232}Th: $\sim2.76\times10^{20}$ & & \multirow{2}{*}{8} & \multirow{2}{*}{1.12} & \cite{ref22} \\
 & & & \textsuperscript{229}Th: $\sim5.4\times10^{17}$ & 9.318 & & & \cite{ref21} \\
LiF & $Fm\bar{3}m$ & 870 & \textsuperscript{232}Th: $\sim8.23\times10^{18}$ & 9.601 & 6 & 0.9 & \cite{ref29} \\
BaLiF\textsubscript{3} & $Pm\bar{3}m$ & 830 & \textsuperscript{232}Th: $\sim3.28\times10^{18}$ & 8.950 & 12 & 1.61 & \cite{ref30} \\
LiSrAlF\textsubscript{6} & $P\bar{3}1c$ & 766 & \textsuperscript{229}Th: $\sim1.0\times10^{17}$ & -- & 6 & 1.32 & \cite{ref73,ref74} \\
MgF\textsubscript{2} & $P4_2/mnm$ & 1261 & -- & 11.6 & 6 & 0.86 & \cite{ref75} \\
SrB\textsubscript{4}O\textsubscript{7} & $Pmn2_1$ & 972 & \textsuperscript{232}Th: $\sim10^{18}$ & 9.695 & 9 & 1.31 & \cite{ref35} \\
\bottomrule
\end{tabular}
}
\end{table}

\section{Conclusion}\label{sec:4}

In this work, pure BaF\textsubscript{2} and \textsuperscript{232}Th-doped BaF\textsubscript{2} crystals were prepared by the Bridgman method, and structural characterization was combined with first-principles calculations to investigate the local defect environments associated with Th doping. The XRD and Laue results support retention of the cubic BaF\textsubscript{2} host structure and single-crystal characteristics in the measured regions of the Th-doped samples. ICP results show that \textsuperscript{232}Th can be successfully incorporated into the BaF\textsubscript{2} crystals, with a relatively high segregation coefficient of 1.38. Raman measurements identify a lower-frequency Raman-active mode in BaF\textsubscript{2} than those reported for CaF\textsubscript{2} and SrF\textsubscript{2}. The Th monomer calculations support the 90° {[}ThBa\textsuperscript{••}-2F\textsubscript{i}′{]} configuration as exhibiting relatively small structural distortion from a kinetic perspective and a low thermodynamic formation energy. The PBE0-calculated band gaps of pure BaF\textsubscript{2} (9.379 eV) and Th:BaF\textsubscript{2} (9.074 eV) correspond to nominal absorption-edge wavelengths of \textasciitilde132.2 nm and 136.6 nm, respectively. Th doping induces a 4.4 nm redshift, while both absorption edges stay below 148.4 nm, supporting further evaluation of BaF\textsubscript{2} as a host for solid-state nuclear optical clocks. Further calculations of 31 charge-neutral candidate Th dimer configurations in fluorite BaF\textsubscript{2} identify six configurations of 1, 3, 9, 13, 15, and 27, with similar Th--Th distances and local coordination features, relatively high structural symmetry, and low cluster binding energies. These results provide clues to defect structures that may exist in actual crystals.

The grown pure BaF\textsubscript{2} crystal exhibits transmittances of 64.43\% at 148.4 nm and 67.15\% at 150 nm, providing a baseline for evaluating VUV losses in the Th-doped system. Future work should correlate the VUV transmission spectra of \textsuperscript{232}Th-doped BaF\textsubscript{2} crystals with oxygen impurities, fluorine nonstoichiometry, and macroscopic scattering defects to identify the main factors limiting transmission near 148.4 nm and guide optimization of the growth process and crystal properties. Process optimization can include drying and appropriate fluorination of the raw materials to reduce sources of water and oxygen, controlling the furnace atmosphere to mitigate impurity contamination and fluorine depletion, adjusting the temperature gradient and cooling process to improve solidification stability, and performing in situ annealing under atmospheres such as Ar or CF\textsubscript{4}. Experimental reports of restored VUV transparency through fluorine replenishment during the post-growth fluorination annealing of \textsuperscript{229}Th:CaF\textsubscript{2} crystals provide a reference for controlling fluorine content. These results provide a structural basis and an experimental starting point for investigating Th:BaF\textsubscript{2} as a candidate host for nuclear clocks.

\section*{Author Contributions}

Lin Li: Conceptualization, methodology, investigation, formal analysis, visualization, writing---original draft. Yuxiang Li: Methodology, formal analysis, investigation, writing---review \& editing. Qiaorui Gong: Methodology, investigation, Shanming Li: Project administration, investigation, Mingzhu He: Methodology, investigation, Shuang Cai: Methodology, formal analysis, Peixiong Zhang: Project administration, resources, supervision, Chengchun Zhao: Supervision, writing---review \& editing, Yin Hang: Resources, supervision, writing---review and editing. Longsheng Ma: Conceptualization, project administration, Shining Zhu: Project administration, resources. (Based on Contributor Role Taxonomy, CRediT)

\section*{Conflict of Interest}

The authors declare no competing financial interest.

\section*{Data Availability Statement}

Data will be made available on request.

\section*{Acknowledgements}

This research was supported by the National Natural Science Foundation of China (NSFC) (12341402, 12341403, U22A20132), Strategic Priority Research Program of the Chinese Academy of Sciences (XDB0920000), and Zhangjiang Laboratory (ZJSP21A001D). The authors gratefully acknowledge the State Key Laboratory of New Textile Materials and Advanced Processing at Wuhan Textile University for its testing assistance.

\end{document}